\documentclass[format=sigconf, screen, natbib=false]{acmart}

\def\myauthors{David Otero, Javier Parapar and Nicola Ferro}
\def\mytitle{How Discriminative Are Your Qrels?
             How To Study the Statistical Significance of Document Adjudication Methods}

\usepackage{multirow}
\usepackage{soul}
\usepackage{subcaption}
\usepackage{amsmath}
\usepackage{graphicx}
\usepackage{url}
\usepackage[inline]{enumitem}
\usepackage{acronym}
\usepackage{booktabs}
\usepackage{longtable}
\usepackage{pdflscape}
\usepackage{tabularx}
\usepackage{afterpage}
\usepackage{siunitx}
\usepackage{algpseudocode, algorithm}
\usepackage{calc}
\usepackage[backend=biber,style=acmnumeric,datamodel=acmdatamodel,backref=true,backreffloats=false,sortcites]{biblatex}
\usepackage{todonotes}
\usepackage{datenumber}
\usepackage{colortbl}
\usepackage{cleveref}
\usepackage{pifont}

\hypersetup{
	pdfauthor={\myauthors},
	pdftitle={\mytitle},
	pdfsubject={Information Retrieval},
}
\crefname{enumi}{}{}
\graphicspath{{plots/}}
\algblock{Input}{EndInput}
\algnotext{EndInput}
\algblock{Output}{EndOutput}
\algnotext{EndOutput}
\newcommand{\Desc}[2]{\State \makebox[2em][l]{#1}#2}
\newcommand{\ra}[1]{\renewcommand{\arraystretch}{#1}}

\newcommand{\un}[1]{\underline{#1}}
\newcommand{\myparagraph}[1]{\paragraph*{\hspace*{-\parindent}\normalsize\bf#1}}
\DefineBibliographyStrings{american}{%
  backrefpage={cited on p.},
  backrefpages={cited on pp.}
}
\acrodef{IR}{Information Retrieval}


\AtBeginDocument{}
\copyrightyear{2023}
\acmYear{2023}
\setcopyright{rightsretained}
\acmConference[CIKM '23]{Proceedings of the 32nd ACM International Conference on Information and Knowledge Management}{October 21--25, 2023}{Birmingham, United Kingdom}
\acmBooktitle{Proceedings of the 32nd ACM International Conference on Information and Knowledge Management (CIKM '23), October 21--25, 2023, Birmingham, United Kingdom}
\acmDOI{10.1145/3583780.3614916}
\acmISBN{979-8-4007-0124-5/23/10}

\makeatletter
\gdef\@copyrightpermission{
  \begin{minipage}{0.3\columnwidth}
    \href{https://creativecommons.org/licenses/by/4.0}{\includegraphics[width=0.90\textwidth]{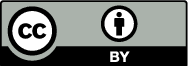}}
  \end{minipage}\hfill
  \begin{minipage}{0.7\columnwidth}
    \href{https://creativecommons.org/licenses/by/4.0/}{This work is licensed under a Creative Commons Attribution International 4.0 License.}
  \end{minipage}
  \vspace{5pt}
}
\makeatother
\author{David Otero}
\email{david.otero.freijeiro@udc.es}
\orcid{0000-0003-1139-0449}
\affiliation{
	\department[0]{Information Retrieval Lab, CITIC}
	\institution{Universidade da Coru\~{n}a}
	\streetaddress{Facultad de Inform\'{a}tica, Campus de Elvi\~{n}a s/n}
	\city{A Coru\~{n}a}
	\country{Spain}
	\postcode{15071}
}

\author{Javier Parapar}
\email{javier.parapar@udc.es}
\orcid{0000-0002-5997-8252}
\affiliation{
	\department[0]{Information Retrieval Lab, CITIC}
	\institution{Universidade da Coru\~{n}a}
	\streetaddress{Facultad de Inform\'{a}tica, Campus de Elvi\~{n}a s/n}
	\city{A Coru\~{n}a}
	\country{Spain}
	\postcode{15071}
}

\author{Nicola Ferro}
\email{ferro@dei.unipd.it}
\orcid{0000-0001-9219-6239}
\affiliation{
	\institution{University of Padua}
	\city{Padova}
	\country{Italy}
}
\begin{document}

%%%%%%%%%%%%%%%%%%%%%%%%%%%%%%%%%%%%%%%%%%%%%%%%%
% Title
%%%%%%%%%%%%%%%%%%%%%%%%%%%%%%%%%%%%%%%%%%%%%%%%%
\title[How To Study the Statistical Significance of Document Adjudication Methods]{\mytitle}

%%%%%%%%%%%%%%%%%%%%%%%%%%%%%%%%%%%%%%%%%%%%%%%%%
% Abstract
%%%%%%%%%%%%%%%%%%%%%%%%%%%%%%%%%%%%%%%%%%%%%%%%%
\begin{abstract}
Creating test collections for offline retrieval evaluation
requires human effort to judge documents' relevance.
This expensive activity motivated much work in developing
methods for constructing benchmarks with fewer assessment costs.
In this respect, adjudication methods actively decide both which documents
and the order in which experts review them,
in order to better exploit the assessment budget or to lower it.
Researchers evaluate the quality of those methods
by measuring the correlation between the known gold ranking of systems under the full collection
and the observed ranking of systems under the lower-cost one.
This traditional analysis ignores whether and how the low-cost judgements
impact on the statistically significant differences among systems
with respect to the full collection.
We fill this void by proposing a novel methodology
to evaluate how the low-cost adjudication methods
preserve the pairwise significant differences between systems as the full collection.
In other terms, while traditional approaches look for stability in answering the question
``is system A better than system B?'',
our proposed approach looks for stability in answering the question
``is system A significantly better than system B?'',
which is the ultimate questions researchers need to answer
to guarantee the generalisability of their results.
Among other results, we found that the best methods in terms of ranking of systems correlation
do not always match those preserving statistical significance.
\end{abstract}

\maketitle

%%%%%%%%%%%%%%%%%%%%%%%%%%%%%%%%%%%%%%%%%%%%%%%%%
% Sections
%%%%%%%%%%%%%%%%%%%%%%%%%%%%%%%%%%%%%%%%%%%%%%%%%
% !TeX spellcheck = en_GB

%%%%%%%%%%%%%%%%%%%%%%%%%%%%%%%%%%%%%%%%%%%%%%%%%
% Introduction
%%%%%%%%%%%%%%%%%%%%%%%%%%%%%%%%%%%%%%%%%%%%%%%%%
\section{Introduction}
\label{sec:intro}

\ac{IR} is a field with a strong focus on evaluation~\cite{Harman2011,Voorhees2001a},
whose main purpose is to empirically measure the effectiveness of retrieval systems.
Offline batch evaluation allows researchers to perform experiments under controlled conditions
and enables the reproducibility of the results.
It is based on test collections, which consist of a corpus of documents,
topics, and relevance judgements (also called assessments, or \textit{qrels})~\cite{Sanderson2010}.
Acquiring the assessments for creating these collections is costly,
since human experts have to judge the documents' content and decide which ones are relevant for each topic.
The advantage is that once the collections are created,
it is straightforward and cheap to conduct as many experiments as needed
to evaluate and compare the performance of (new) \ac{IR} systems~\cite{Voorhees2005}.

The first and small test collections had complete judgements~\cite{Harman2011},
containing a human assessment for each topic-document pair,
thus representing the ideal situation in terms of evaluation quality.
However, that exhaustive procedure is only feasible
for collections with a very small corpus.
Nonetheless, small corpora are not the conditions that operational search systems face.
As a consequence, when larger collections arose,
there was the need to implement some kind of \emph{sampling}
so that assessors would not have to judge the relevance of each document for each topic.
However, simple random sampling, the most immediate approach, would not work,
since the number of relevant documents for a topic is extremely small
compared to the size of the corpus of documents.
Thus, a random sample would end up consisting of (almost all) non-relevant documents.
The first solution to this problem was the \textit{pooling} technique
implemented by TREC~\cite{Sparck1975,Voorhees2005}.
With this technique, assessors only judge a subset of the corpus, the \emph{pool}.
For each topic, the pool consists of the union of the top-$k$ documents
retrieved by several search systems for that topic.
The assessors judge the relevance of the documents in the pool while the rest,
i.e. the non-pooled documents, are assumed to be non-relevant.
Top-$k$ pooling builds on the assumption that \ac{IR} systems
try to push relevant documents towards the top of the ranking
and thus there is a good chance to pool most of the relevant documents for a topic,
provided that $k$ is deep enough and the pooled systems are diverse enough.
However, the number of judgements that an assessor can perform,
i.e. the \textbf{budget}, is limited and, therefore,
there is a trade-off with the depth $k$ of the pool and the number of pooled systems,
since the more they grow, the higher the number of documents in the pool.

Pooling does not guarantee finding all the relevant documents for a topic but,
as said, it strives to find a very good share of them.
Researchers are interested in comparing systems in order to answer the fundamental question
``is system A better than system B?''.
Answering this question requires a good estimate of system performance
rather than absolute performance scores which, in turn, would demand finding all the relevant documents.
Therefore, the quality of a pool is \emph{traditionally} measured on its ability to
\emph{fairly rank} systems, i.e. to fairly compare them.
This is not limited to the systems which were actually pooled,
but it should also hold for systems which were not pooled~\cite{Zobel1998},
to ensure the future reusability of a test collection also with new systems.

However, collections kept growing in size, and just judging deep pools over a diverse set of systems
stopped to be a practicable approach as well~\cite{Voorhees2022b}.
Therefore, much work has focused on developing alternative methods
to better select which documents to pool and judge by performing some sort of \emph{focused sampling},
aimed at picking documents which more probably turn out to be relevant and
better employing the assessor budget
or allowing for lower budgets at a comparable quality~\cite{Losada2017a,Rahman2019}.
A method that \emph{actively} decides which document to judge next is called an \textbf{adjudication method}.
However, alternative prioritisation models may introduce biases or incompleteness in the judgements,
hampering the future reusability of an test collection~\cite{Voorhees2018}.

Therefore, the quality of new adjudication methods is \emph{traditionally} assessed
by checking that they rank systems as closely as possible to the full set of judgements of a (good quality) top-$k$ pool,
ensuring that they can still properly answer the question ``is system A better than system B?''.
This is quantified by computing the correlation, e.g. Kendall's $\tau$~\cite{Kendall1938,Kendall1948},
between the ranking of systems produced by an adjudication method and by the full top-$k$ pool.
The rationale is that if this correlation is high, 
one may assume the validity of the new method and aim to use it in the future for building
new test collections at a comparable quality but with a lower assessment cost.

However, the question researchers are really interested in is rather
``is system A \emph{statistically significantly} better than system B?'',
since this ensures that observed differences are not due just to the randomness
present in the construction process of a collection and, especially,
that the found differences would \emph{generalise} better and
still hold in operational settings~\cite{Fuhr2018,Sakai2021}.
The problem is that the above correlation measures ignore whether the evaluated systems'
statistical significance is preserved.

Let us better explain this problem with an example.
Let us assume we have three different \ac{IR} systems,
\texttt{Sys1}, \texttt{Sys2} and \texttt{Sys3},
and that their true ranking, given by the full top-$k$ pool, is
(\texttt{Sys1}, \texttt{Sys2}, \texttt{Sys3}).
We perform a significance test between all possible pairwise comparisons
and we obtain that \texttt{Sys1} is significantly better than \texttt{Sys2}
and \texttt{Sys3}, and \texttt{Sys2} is also significantly better than \texttt{Sys3}.
Then, we create a new set of judgements using some adjudication method and repeat the above procedure.
Using this new pool, we find the same ranking of systems as when using the full top-$k$ pool,
leading to a perfect correlation and concluding that the adjudication method is fully equivalent,
but less costly, than the full top-$k$ pool.
However, we do not know anything about the significance between systems.
If we repeat the same significance test using the new pool instead,
we may not find any significant difference between any pair.
We may thus conclude that there is no evidence of any system being different from the rest.
This would be the opposite conclusion than the one drawn on the full top-$k$ pool,
where all the system pairs were significantly different.

In this work, our objectives are two-fold.
First, we aim to propose a new approach to evaluate the validity of low-cost adjudication methods,
focusing on how they preserve the statistically significant differences between systems.
Second, we analyse some state-of-the-art adjudication methods using our new approach to gain new insights about them.
In particular, we aim to answer the following research questions:
\begin{enumerate*}[label=\textbf{RQ\arabic*}]
  \item \label{rq1}
  Are the adjudication methods able to preserve the same statistically significant differences
  as the full top-$k$ pool?
  
  \item \label{rq2}
  When adjudication methods fail to see a real significant difference,
  do they follow any distinguishable pattern in terms of system position in the ranking?
  
  \item \label{rq3}
  Are the adjudication methods able to preserve the same statistically significant differences as
  the full top-$k$ pool for new (non-pooled) systems?
\end{enumerate*}

The rest of the paper is organised as follows:
\Cref{sec:rw} introduces past work;
\Cref{sec:method} explains our methodology;
\Cref{sec:experiments} and~\Cref{sec:results} report our experiments;
and, finally, \Cref{sec:fw} draws conclusions and presents some ideas for future work.
% !TeX spellcheck = en_GB

%%%%%%%%%%%%%%%%%%%%%%%%%%%%%%%%%%%%%%%%%%%%%%%%%
% Related Work
%%%%%%%%%%%%%%%%%%%%%%%%%%%%%%%%%%%%%%%%%%%%%%%%%
\section{Related Work}
\label{sec:rw}

How to build high-quality experimental collections for retrieval evaluation
is still an open research question~\cite{Craswell2021a,Voorhees2022b,Voorhees2022a}.
Research in adjudication methods looks for ways of prioritising the pooled documents
so that the assessors expend their effort in judging relevant documents.
In this way, we may only need to judge some of the pooled documents
while maintaining the quality of the judgements,
thus making more efficient use of the resources.

Losada et al.~\cite{Losada2017a} proposed a series of sampling methods
based on the multi-armed bandit problem.
The multi-armed bandit problem~\cite[Chapter~2]{Sutton2018} has been a subject of research for decades
in Reinforcement Learning (RL), statistics and other fields.
These methods bring ideas from RL to the task of document adjudication for building test collections.
They apply Bayesian principles to this problem,
formalising the uncertainty associated with reviewing a document from a pooled system.
Other works have also explored the development of adjudication methods~\cite{Cormack1998,Li2017,Rahman2019,Moffat2007}
Section~\ref{sec:experiments} provides further details about the state-of-the-art adjudication methods
under experimentation.
Adjudication methods have shown remarkable improvements in bringing relevant documents
earlier in the pooling process,
and indeed they were used to build the collection of the TREC Common Core Track of 2017~\cite{Allan2017}.
However, the quality of the judgements produced with a limited budget is
still an open question~\cite{Voorhees2018}.

Previous work on adjudicating methods used a series of metrics to evaluate the quality of these algorithms.
The commonest is Kendall's~$\tau$~\cite{Kendall1938,Kendall1948} correlation,
which researchers use to measure how well a new adjudication method can induce the gold ranking of systems,
i.e. the one on the full top-$k$ pool.
Another top-weighted correlation, $\tau_{AP}$~\cite{Yilmaz2008}, is also common.
This correlation penalises swaps in higher positions more.
In some works~\cite{Voorhees2018,Voorhees2022b},
they also measure the change in the ranking position of the system that suffers the highest drop
as a measure of the reusability of an experimental collection.
The problem with all these measures, as we already introduced earlier,
is that they ignore the significance between the scores of the systems.
If we ignore this, it is meaningless to account for ranking swaps.

In this work, we propose a new methodology to evaluate low-cost adjudication methods that,
instead of focusing only on the ranking of the systems,
focuses on evaluating how well a method preserves the real pairwise significant differences.

Statistical significance testing is of paramount importance in \ac{IR},
and studying the properties of significance tests is an active area of research~
\cite{Banks1999,Carterette2017,Carterette2012,Cormack2006,Cormack2007b,Ferro2019,Ferro2022,Hull1993,Sakai2016b,Sanderson2005,Savoy1997,Webber2008,Parapar2021,Urbano2019,Parapar2020,Urbano2013}.
However, this is out-of-scope for the present work which, instead,
focuses on considering the output of a statistical significance test
as a way to assess the quality of an adjudication method.
% !TeX spellcheck = en_GB

%%%%%%%%%%%%%%%%%%%%%%%%%%%%%%%%%%%%%%%%%%%%%%%%%
% Method
%%%%%%%%%%%%%%%%%%%%%%%%%%%%%%%%%%%%%%%%%%%%%%%%%
\section{Method}
\label{sec:method}

Let $S = \{ s_i \}$, $\lvert S \rvert = n$, be the set of systems under experimentation, and 
let $G$ be the \emph{gold assessments} (also said gold qrels), i.e. the full top-$k$ pool.
Using an effectiveness measure of choice,
we compute the per-topic scores for each of the $n$ systems and
we perform a statistical test for each pairwise comparison between systems.
From this test, we obtain, for each pair of systems $s_i$ and $s_j$ ($i < j \leq n$),
a triplet $\langle s_i, s_j, c \rangle$,
where $c \in \{>, \gg, <, \ll\}$, denoting the four outcomes we are interested in:
$s_i$ is better than $s_j$ ($s_i > s_j$),
$s_i$ is \emph{significantly} better than $s_j$ ($s_i \gg s_j$),
$s_j$ is  better than $s_i$ ($s_i < s_j$), or
$s_j$ is \emph{significantly} better than $s_i$ ($s_i \ll s_j$).

Now we use $R_G$ to denote the set of triplets that result from the statistical test
performed using the gold qrels.
Similarly, we use $L$ to denote the qrels obtained with a low-cost adjudication method
($L \subseteq G$) and
$R_L$ to denote the set of triplets that result from the statistical test performed with them.
Note that $\lvert R_G \rvert = \lvert R_L \rvert = \frac{n (n - 1)}{2}$.
Finally, we use $T_G$ to denote the set of comparisons from $R_G$ that are significantly different,
that is, the set triplets for which $c \in \{\ll, \gg\}$,
and $T_L$ for the significantly different comparisons obtained with the low-cost assessments.

As we already explained, we are interested in studying to what extent
the judgements produced by different low-cost adjudication methods preserve the statistically
significant differences between systems we observe when using the gold qrels.
The idea here is that if the low-cost method is able to preserve such differences,
we could confidently use it to build new collections in the future with fewer assessment costs.
Thus, we compare how $T_G$ and $T_L$ agree with each other
using the measures described in the following section.

\subsection{Measures}

%%%%%%%%%%%%%%%%%%%%%%%%%%%%%%%%%%%%%%%%%%%%%%%%%
% Kendall's tau
%%%%%%%%%%%%%%%%%%%%%%%%%%%%%%%%%%%%%%%%%%%%%%%%%
\myparagraph{Kendall's $\tau$}

Kendall's $\tau$ is the measure traditionally used to evaluate adjudication methods.
It computes the correlation between the ranking of systems under the gold qrels setting and
the one under the qrels produced with the different adjudication methods.

Given two rankings over the same set of items,
Kendall's~$\tau$ computes how many items are swapped as follows: $\tau = (P - Q)/\binom{n}{2}$,
where $P$ is the number of concordant pairs
(pairs of systems ranked in the same relative order in both lists),
$Q$ is the number of discordant pairs (swapped pairs of systems),
and $\binom{n}{2} = \frac{n (n - 1)}{2}$ is the number of total pairs, given that we have $n$ items.

%%%%%%%%%%%%%%%%%%%%%%%%%%%%%%%%%%%%%%%%%%%%%%%%%
% Precision and Recall
%%%%%%%%%%%%%%%%%%%%%%%%%%%%%%%%%%%%%%%%%%%%%%%%%
\myparagraph{Precision and Recall}

We consider the \emph{Precision} (P) and \emph{Recall} (R) of the significantly different pairs
detected by the low-cost adjudication methods, defined as follows:

\begin{equation*}
  P = \frac{\lvert T_G \cap T_L \rvert}{\lvert T_L \rvert},~
  R = \frac{\lvert T_G \cap T_L \rvert}{\lvert T_G \rvert}
\end{equation*}

\noindent where $\lvert T_G \cap T_L \rvert$ is the number of significantly different pairs
common to both the gold and adjudication qrles,
i.e. the correct ones when assuming the gold qrels detect the ``true'' differences.
Precision indicates how much ``noise'' is introduced by an adjudication method,
meant as additional significant differences not detected by gold qrels;
Recall indicates how many of the total possible significant differences
are not detected by an adjudication method.

%%%%%%%%%%%%%%%%%%%%%%%%%%%%%%%%%%%%%%%%%%%%%%%%%
% Agreements
%%%%%%%%%%%%%%%%%%%%%%%%%%%%%%%%%%%%%%%%%%%%%%%%%
\myparagraph{Agreements.}

We consider an adaptation of a series of agreement measures that have been used in past work~\cite{Ferro2022,Moffat2012,Urbano2013,Faggioli2021b}.
Note that, while Kendall's $\tau$ and Precision/Recall focus on ranking of systems (the former) or
on matching significantly different pairs (the latter) in isolation,
the following agreement measures consider them jointly.

\begin{itemize}[wide, labelindent=0pt, itemsep=2pt]
  \item
  \textbf{Active Agreements (AA)}: the set of consistent outcomes between both methods.
  This is,
  $\langle s_i, s_j, \gg \rangle \in T_G$ and $\langle s_i, s_j, \gg \rangle \in T_L$ or
  $\langle s_i, s_j, \ll \rangle \in T_G$ and $\langle s_i, s_j, \ll \rangle \in T_L$.
  This is the best possible case, and thus, the larger AA are, the better.

  \item \textbf{Active Disagreements (AD)}: the set of opposite outputs between both methods.
  This is,
  $\langle s_i, s_j, \gg \rangle \in T_G$ and $\langle s_i, s_j, \ll \rangle \in T_L$, or
  $\langle s_i, s_j, \ll \rangle \in T_G$ and $\langle s_i, s_j, \gg \rangle \in T_L$.
  This is the worst possible case, since it means that both methods reach complete opposite
  conclusions for a given pair. Thus, the lesser, the better.

  \item 
  \textbf{Mixed Agreements (MA)}:
  we have four possible options:
  \ding{202}
  $\langle s_i, s_j, \ll \rangle \in T_G$ and $\langle s_i, s_j, < \rangle \in T_L$, or
  \ding{203}
  $\langle s_i, s_j, \gg \rangle \in T_G$ and $\langle s_i, s_j, > \rangle \in T_L$, or
  \ding{204}\label{204}
  $\langle s_i, s_j, < \rangle \in T_G$ and $\langle s_i, s_j, \ll \rangle \in T_L$, or
  \ding{205}
  $\langle s_i, s_j, > \rangle \in T_G$ and $\langle s_i, s_j, \gg \rangle \in T_L$.
  We distinguish between MA\textsubscript{G} (\ding{202} and \ding{203}),
  which counts the cases where the
  adjudication method was not able to see a gold significant difference.
  Conversely, MA\textsubscript{L} (\ding{204} and \ding{205}) counts the cases where a 
  low-cost method sees a significant difference that is not in the gold qrels.
  Note that MA\textsubscript{G}~+~MA\textsubscript{L}~=~MA

  \item 
  \textbf{Mixed Disagreements (MD)}:
  we also have four possible cases here:
  \ding{206}
  $\langle s_i, s_j, \ll \rangle \in T_G$ and $\langle s_i, s_j,   > \rangle \in T_L$, or
  \ding{207}
  $\langle s_i, s_j, \gg \rangle \in T_G$ and $\langle s_i, s_j,   < \rangle \in T_L$, or
  \ding{208}
  $\langle s_i, s_j,   > \rangle \in T_G$ and $\langle s_i, s_j, \ll \rangle \in T_L$, or
  \ding{209}
  $\langle s_i, s_j,   < \rangle \in T_G$ and $\langle s_i, s_j, \gg \rangle \in T_L$.
  Here, as with MA, we also distinguish between MD\textsubscript{G} (\ding{206} and \ding{207})
  and MD\textsubscript{L} (\ding{208} and \ding{209})
\end{itemize}

\myparagraph{Bias.}
Analogously to Ferro and Sanderson~\cite{Ferro2022}, we also consider the \emph{publication bias},
i.e. the likelihood of a researcher publishing a significant result using an adjudication method
when in fact a significance test on the gold qrels would have produced either no significance
(MA, MD) or a significant result in the opposite direction (AD). We define it as follows:

\begin{equation*}
  Bias = 1 - \frac{AA}{AA + AD + MA_{L} + MD_{L}}
\end{equation*}

A value of 0\% means that every significance detected by an adjudication method
leads to the same conclusions (and publication) as those of the gold qrels.
Conversely, a value of 100\% means that every significance detected by an adjudication method
leads to opposite conclusions (and publication) to those of the gold qrels.
Thus, the lower the bias, the better.
Note that, differently from Ferro and Sanderson~\cite{Ferro2022},
we do not consider the whole MA and MD but just MA\textsubscript{L} and MD\textsubscript{L},
since we are interested only in the publication bias induced by the adjudication method.
This metric tries to measure the situations where a researcher sees a significant outcome 
under the reduced pools when, in reality, it would be a different conclusion under the gold qrels.

%%%%%%%%%%%%%%%%%%%%%%%%%%%%%%%%%%%%%%%%%%%%%%%%%
% FWER
%%%%%%%%%%%%%%%%%%%%%%%%%%%%%%%%%%%%%%%%%%%%%%%%%
\subsection{Family-Wise Error Rate (FWER)}

Performing \emph{multiple comparisons}---in our case between each pair of systems---leads to an increase of the \emph{Type I error},
i.e. incorrectly rejecting the null hypothesis, and inflates the number of significant differences found~\cite{Hochberg1987,Hsu1996,Sakai2018}.

The Type I error probability is equal to the significance level $\alpha$ and,
as the number of comparisons increases, this probability also does.
If we perform $k$ different system comparisons,
the probability of correctly accepting the null hypothesis for all of them
is equal to $(1 - \alpha)^k$.
Thus, the probability of committing at least one Type I error is $1 - (1 - \alpha)^k$.
This is the \textit{family-wise error rate} (FWER).
If we have, for example, $\alpha = 0.05$ and $k = 6$ comparisons (4 systems, $\frac{4(4-1)}{2}=6$),
this probability would rise to $0.264$, which is not acceptable.
For this reason, when we perform multiple comparisons, we should employ a technique to adjust the p-values,
so that the FWER stays below $\alpha$.
Obviously, this has the side-effect of reducing the \emph{power} of the statistical test and increasing the number of \emph{Type II errors},
i.e. not detecting an actual significant difference.

There are several options to control the FWER in a multiple comparison situation.
The Bonferroni correction, for example, is a post-hoc correction
where, if we have $k$ different comparisons, we should use $p < \frac{\alpha}{k}$ as our significance level in each pairwise comparison.
However, the Bonferroni correction is known to be too conservative and to reduce the power of a test too much,
especially when the number of comparisons increases as in our case.
Therefore, we employ the randomised version of the Tukey Honestly Significant Difference (HSD) test~\cite{Carterette2012,Sakai2018}.
This is a nonparametric computer-based generalisation of the common permutation test
for handling more than 2 systems.
At each permutation, the test perturbs the array of system scores of each topic, and, after this perturbation,
computes the difference between the maximum and minimum average system scores.
Then the test counts how many times the actual differences between system average performance is greater than the permuted mean 
to determine if it is \textit{honestly} significant~\cite{Carterette2012}.
The Tukey HSD test produces a p-value for each pairwise comparison,
which can be compared to the significance level $\alpha$ to decide whether that pair of systems is significantly different or not.
Algorithm~\ref{alg:tukey} (adapted from prior work~\cite{Carterette2012,Sakai2018}) shows the details of our implementation.

%%%%%%%%%%%%%%%%%%%%%%%%%%%%%%%%%%%%%%%%%%%%%%%%%
% Tukey pseudocode
%%%%%%%%%%%%%%%%%%%%%%%%%%%%%%%%%%%%%%%%%%%%%%%%%
\begin{algorithm}
  \caption{Paired Randomised Tukey HSD}
  \label{alg:tukey}
  \begin{algorithmic}
    \Input
      \Desc{$X$}{$m\times n$ topic-system scores matrix.}
      \Desc{$B$}{number of permutations.}
    \EndInput
    \Output
      \Desc{$P$}{$n\times n$ matrix holding a p-value for each pairwise system comparison.}
    \EndOutput
    \For{$k \gets 1$ to $B$}
      \State initialise $m \times n$ matrix $X'$
      \For{each topic $t$}
        \State $\textnormal{row t of}~X' \gets \textnormal{permutation of values in row t of}~X$
      \EndFor
      \State $d' \gets \max_{i}\bar{X'_i} - \min_{j}\bar{X'_j}$ \Comment{$\bar{X'_i}$ is the mean of column $i$}
      \For{each pair of systems $i, j$}
        \If{$d' > \lvert \bar{X_i} - \bar{X_j} \rvert$}
          \State $P_{i,j} \gets P_{i,j} + \frac{1}{B}$
        \EndIf
      \EndFor
    \EndFor
  \end{algorithmic}
\end{algorithm}
% !TeX spellcheck = en_GB

\section{Experimental Setup}
\label{sec:experiments}

\myparagraph{Collections}

We employ the TREC-8 ad hoc collection,
known to have a very high-quality pool~\cite{Voorhees2000,Voorhees2022a}.
It includes 129 system submissions, retrieving \num{1000} documents for each topic, and 50 topics.
Official relevance judgements are based on a pool of depth 100 over 71 out of 129 submitted runs,
resulting in \num{86830} assessments across all 50 topics.
The average pool size per topic is \num{1736}, while the maximum and the minimum are \num{2992} and \num{1046}, respectively.
Additionally, we use the collection from the document ranking task of TREC 2021 Deep Learning track~\cite{Craswell2021b},
which adopted a shallow pooling approach at depth 10, then enlarged with a method based on active learning.
We used only the documents in the top-10 pools as our gold qrels to provide a fairer comparison to the case of TREC-8.
It includes 66 runs, retrieving 100 documents for each topic,
and \num{13058} judgements made by NIST assessors over 57 different topics.
The depth-10 pools we used include \num{6510} judgements, with an average pool size of \num{114},
a maximum of \num{226} and a minimum of \num{50}.

\myparagraph{Adjudication methods}

We consider a series of state-of-the-art adjudication methods.

\begin{itemize}[wide, labelindent=0pt, itemsep=1.5pt]
  \item \textbf{top-$k$ pooling.}
  We adapt the standard method used in TREC to limited-budget situations.
  When limiting the budget of assessments, we choose a $k$ deep enough to fill that budget.
  Then, pooled documents are sorted by their document identifier~\cite{Voorhees2005}.

  \item \textbf{MoveToFront (MTF).}
  MTF is a dynamic adjudication method proposed by Cormack and colleagues~\cite{Cormack1998}
  that has been acknowledged as a robust adjudication method~\cite{Altun2020}.

  \item \textbf{MaxMean (MM), MM Non Stationary (MM-NS), Thompson Sampling (TS) and
  TS Non Stationary (TS-NS).}
  Bandit-based methods for document adjudication
  apply bayesian principles to formalise the
  uncertainty associated with the probabilities of pulling a positive reward (a relevant document)
  from playing a bandit~\cite{Losada2016a}.

  \item \textbf{Hedge.}
  Hedge is an online learning algorithm adapted for pooling in~\cite{Aslam2003}.
  A more detailed explanation of applying Hedge for pooling can be found in this article~\cite{Losada2017a}.

  \item \textbf{NTCIR top-$k$ prioritization.}
  Documents in the pool are sorted by the number of runs that contain
  the document at or above the depth $k$ (the higher the better),
  ties are solved with the sum of the ranks of that document within the runs (the lower the better)~\cite{Sakai2008}.
\end{itemize}

\myparagraph{Other Settings}

We used Average Precision (AP)~\cite{Buckley2005} and Normalized Discounted Cumulative Gain (NDCG)~\cite{Jarvelin2002} as performance measures to score runs.
We used $\alpha = \num{0.05}$ as significance level and $B = \num{1000000}$ permutations in Tukey HSD test.
Finally, since MTF, MM, MM-NS, TS, and TS-NS have a stochastic nature,
the reported results for those methods are averaged over 50 executions of each.

To ease the reproducibility of the experiments, we release the source code.\footnote{\url{https://github.com/davidoterof/cikm2023}}

% !TeX spellcheck = en_GB

\section{Results and Discussion}
\label{sec:results}

% !TeX spellcheck = en_GB

\begin{table}[b]
    \centering
    \footnotesize
    \caption{Kendall's $\tau$, Precision and Recall (see \Cref{sec:method}) of each
    adjudication method for a varying number of judgements per topic. 100 and 300 are the budget of judgements
    per topic. Parentheses indicate the size of this budget with respect to the full pool.  We used the 71 pooled systems of TREC-8.
    For each column, best values are \textbf{bolded} and worst ones \un{underlined}.}
    \label{tab:pyr-trec8-pooled}
    \setlength\tabcolsep{2pt}
    \begin{tabular}{@{}
                    l@{\hspace{1\tabcolsep}}
                    r@{\hspace{0.8\tabcolsep}}
                    r@{\hspace{0.8\tabcolsep}}
                    r@{\hspace{1\tabcolsep}}
                    c@{\hspace{1\tabcolsep}}
                    r@{\hspace{0.8\tabcolsep}}
                    r@{\hspace{0.8\tabcolsep}}
                    r@{\hspace{1\tabcolsep}}
                    c@{\hspace{1\tabcolsep}}
                    r@{\hspace{0.8\tabcolsep}}
                    r@{\hspace{0.8\tabcolsep}}
                    r@{\hspace{1\tabcolsep}}
                    c@{\hspace{1\tabcolsep}}
                    r@{\hspace{0.8\tabcolsep}}
                    r@{\hspace{0.8\tabcolsep}}
                    r@{}}
        \toprule  
        \multirow{2.3}{*}{\textbf{Method}} &\multicolumn{3}{c}{\textbf{MAP/100} (6\%)}&&\multicolumn{3}{c}{\textbf{MAP/300} (17\%)}&&\multicolumn{3}{c}{\textbf{NDCG/100} (6\%)}&&\multicolumn{3}{c}{\textbf{NDCG/300} (17\%)}\\
        \cmidrule{2-4}\cmidrule{6-8}\cmidrule{10-12}\cmidrule{14-16}
        & \multicolumn{1}{c}{\textbf{$\tau$}} & \multicolumn{1}{c}{\textbf{P}} & \multicolumn{1}{c}{\textbf{R}} &&
          \multicolumn{1}{c}{\textbf{$\tau$}} & \multicolumn{1}{c}{\textbf{P}} & \multicolumn{1}{c}{\textbf{R}} &&
          \multicolumn{1}{c}{\textbf{$\tau$}} & \multicolumn{1}{c}{\textbf{P}} & \multicolumn{1}{c}{\textbf{R}} &&
          \multicolumn{1}{c}{\textbf{$\tau$}} & \multicolumn{1}{c}{\textbf{P}} & \multicolumn{1}{c}{\textbf{R}} \\
        \midrule
        top-$k$    & 0.91 & 0.932 & 0.888 && \un{0.95} & 0.955 & 0.955 && 0.90 & \textbf{0.975} & \un{0.929} && 0.94 & 0.985 & \un{0.970} \\
        MTF        & 0.94 & 0.946 & \textbf{0.961} && 0.97 & 0.962 & 0.980 && 0.91 & \textbf{0.975} & 0.953 && \textbf{0.96} & 0.982 & 0.985 \\
        MM         & \textbf{0.95} & 0.948 & 0.958 && \textbf{0.98} & \textbf{0.969} & \textbf{0.992} && \textbf{0.92} & 0.942 & 0.973 && \textbf{0.96} & 0.976 & \textbf{0.991} \\
        MM-NS      & 0.93 & 0.942 & 0.957 && 0.97 & 0.967 & 0.987 && 0.90 & 0.970 & 0.962 && \textbf{0.96} & \textbf{0.986} & \textbf{0.991} \\
        TS         & \textbf{0.95} & 0.947 & 0.954 && \textbf{0.98} & \textbf{0.969} & 0.991 && \textbf{0.92} & \un{0.940} & 0.970 && \textbf{0.96} & 0.975 & 0.990 \\
        TS-NS      & 0.93 & 0.945 & 0.949 && 0.97 & 0.966 & 0.983 && 0.90 & 0.971 & 0.960 && \textbf{0.96} & 0.985 & \textbf{0.991} \\
        Hedge      & 0.94 & \textbf{0.955} & 0.947 && \textbf{0.98} & 0.968 & 0.980 && \textbf{0.91} & 0.959 & \textbf{0.978} && 0.95 & \un{0.972} & 0.989 \\
        NTCIR      & \un{0.83} & \un{0.900} & \un{0.876} && 0.96 & \un{0.942} & \un{0.925} && \un{0.81} & 0.961 & 0.942 && \un{0.93} & 0.977 & 0.988 \\
        \bottomrule
    \end{tabular}
\end{table}

%%%%%%%%%%%%%%%%%%%%%%%%%%%%%%%%%%%%%%%%%%%%%%%%%
% RQ1
%%%%%%%%%%%%%%%%%%%%%%%%%%%%%%%%%%%%%%%%%%%%%%%%%
\subsection{\cref{rq1}: Preservation of significant differences}

In \Cref{tab:pyr-trec8-pooled}, we report the Kendall's $\tau$, Precision and Recall,
as defined in \Cref{sec:method}, that each adjudication method achieves, while varying the number of assessments per  topic.
We report the scores for 100 judgements per topic (which is a 6\% budget of the original pool), and 300 (17\%).
All this values were obtained using the pooled systems of the TREC-8 collection, which includes 71 different systems.

Regarding Kendall's $\tau$ and consistently with previous findings in the literature,
we see almost every method achieves a very high correlation ($\tau > 0.90$) already at a 6\% of the original budget.
While this means that every method obtains a ranking of systems very similar to the one of the gold qrels,
it also makes it very difficult to distinguish among methods.
Moreover, we can observe that top-$k$ and NTCIR methods stay behind the rest,
leaving room for improvement in developing more efficient adjudication strategies
for building new collections in evaluation workshops.

As we mentioned earlier, Kendall's~$\tau$ does not allow us to know
whether the compared algorithms preserve the same statistically significant differences as the gold qrels.
Therefore, we study to which extent this effect might hold by using the Precision and Recall measures previously introduced.

We observe that every method obtains Precision and Recall values over 90\% in almost all the cases, which is a quite solid result.
Moreover, every method is able to mostly preserve the same differences just having a 6\% of the original budget.
With 300 assessment per topic (17\% of the budget), Recall is (almost) 1.00 for most of the methods,
indicating that they are able to detect all the significant differences of the gold qrels at less than one third of the cost.

It is also interesting to observe that most of them detect some differences that there were not detected in the gold qrels.
Indeed, Precision is lower than 1.00 while Recall is almost 1.00 (all the differences in the gold qrels detected).
In other terms, $T_L$ (the set of significant differences detected by the adjudication method)
is not a proper subset of $T_G$ (the set of  significant differences detected by the gold qrels).
A possible explanation might be that, since reduced pools lack some relevant documents,
the performance difference of some pair of systems (delta AP/NDCG between the two systems in our case)
turns out to be increased with respect to the gold qrels
and this makes the pair significantly different on the reduced pool but not on the gold qrels.
Since more evaluation on this issue would need more experimentation, due to space restrictions we leave this
investigation for future work.

\begin{table}
  \centering
  \footnotesize
  \caption{Relevants, agreements and bias of each adjudication method for a varying
  number of judgements per topic. Parentheses indicate the size with respect to the full pool.
  We used the 71 pooled systems of TREC-8.
  The top-100 full pool includes 4728 relevant documents.
  There are 2485 pairwise comparisons, of which 966 are significant under the gold qrels with MAP (upper half), and 917 with NDCG (lower half).
  For each budget and metric, the best values are \textbf{bolded} and the worst ones are \un{underlined}.}
  \label{tab:agr-trec8-pooled}
  \setlength\tabcolsep{2pt}
  \begin{tabular}{@{}llllrrrrrrrr@{}}
    \toprule
    &&\multirow{2.3}{*}{\textbf{Metric}} && \multicolumn{8}{c}{\textbf{Adjudication method}} \\
    \cmidrule{5-12}
    &&&&\multicolumn{1}{c}{top-\textit{k}}&\multicolumn{1}{c}{MTF}&\multicolumn{1}{c}{MM}&\multicolumn{1}{c}{MM-NS}&\multicolumn{1}{c}{TS}&\multicolumn{1}{c}{TS-NS}&\multicolumn{1}{c}{Hedge}&\multicolumn{1}{c}{NTCIR}\\
    \midrule
    \multirow{20.5}{*}{\rotatebox{90}{\textbf{MAP (966 gold significantly different pairs)}}}&\multirow{10}{*}{\rotatebox{90}{\textbf{Budget per topic: 100 (6\%)}}} &\# rels. && \un{1077} & 1685 & 2148 & 1553 & 2102 & 1514 & \textbf{2170} & 1481 \\
    && AA && 858 & \textbf{929} & 926 & 925 & 922 & 917 & 915 & \un{846} \\
    && MA\textsubscript{total} && 170 & \textbf{90} & \textbf{90} & 98 & 94 & 102 & 94 & \un{185} \\
    && MA\textsubscript{G} && \un{108} & \textbf{37} & 40 & 41 & 44 & 49 & 51 & 91 \\
    && MA\textsubscript{L} && 62 & 52 & 50 & 57 & 50 & 53 & \textbf{43} & \un{94} \\
    && MD\textsubscript{total} && \textbf{0} & \textbf{0} & \un{1} & \textbf{0} & \un{1} & \textbf{0} & \textbf{0}  & 29 \\
    && MD\textsubscript{G} && \textbf{0} & \textbf{0} & \textbf{0} & \textbf{0} & \textbf{0} & \textbf{0} & \textbf{0} & \un{29}\\
    && MD\textsubscript{L} && \textbf{0} & \textbf{0} & \un{1} & \textbf{0} & \un{1} & \textbf{0} & \textbf{0} & \textbf{0} \\
    && AD && 0 & 0 & 0 & 0 & 0 & 0 & 0 & 0 \\
    && Bias && 7\% & 5\% & 5\% & 6\% & 5\% & 5\% & \textbf{4\%} & \un{10\%} \\
    \arrayrulecolor{black!30}
    \cmidrule{3-12}
    \arrayrulecolor{black}
    &\multirow{10}{*}{\rotatebox{90}{\textbf{Budget per top.: 300 (17\%)}}} &\# rels. && \un{2042} & 2923 & \textbf{3628} & 2913 & 3607 & 2868 & 3609 & 2723\\
    && AA && 923 & \textbf{961} & 959 & 954 & 958 & 950 & 947 & \un{894}\\
    && MA\textsubscript{total} && 86 & 43 & \textbf{38} & 44 & 39 & 50 & 50 & \un{127}\\
    && MA\textsubscript{G} && 43 & \textbf{5} & 7 & 12 & 8 & 16 & 19 & \un{72}\\
    && MA\textsubscript{L} && 43 & 38 & \textbf{30} & 32 & \textbf{30} & 33 & 31 & \un{55}\\
    && MD\textsubscript{total} && 0 & 0 & 0 & 0 & 0 & 0 & 0 & 0\\
    && MD\textsubscript{G} && 0 & 0 & 0 & 0 & 0 & 0 & 0 & 0\\
    && MD\textsubscript{L} && 0 & 0 & 0 & 0 & 0 & 0 & 0 & 0\\
    && AD && 0 & 0 & 0 & 0 & 0 & 0 & 0& 0 \\
    && Bias && 4\% & 4\% & \textbf{3\%} & \textbf{3\%} & \textbf{3\%} & \textbf{3\%} & \textbf{3\%} & \un{6\%} \\
    \midrule
    \multirow{20.5}{*}{\rotatebox{90}{\textbf{NDCG (917 gold significantly different pairs)}}}&\multirow{10}{*}{\rotatebox{90}{\textbf{Budget per topic: 100 (6\%)}}} &\# rels. && \un{1077} & 1685 & 2148 & 1553 & 2102 & 1514 & \textbf{2170}& 1481\\
    && AA && \un{852} & 874 & 893 & 883 & 890 & 881 & \textbf{897} & 864 \\
    && MA\textsubscript{total} && 86 & 65 & 79 & 61 & 83 & 62 & \textbf{58} & \un{88} \\
    && MA\textsubscript{G} && \un{65} & 43 & 24 & 34 & 27 & 36 & \textbf{20} & 53 \\
    && MA\textsubscript{L} && \textbf{21} & 22 & 55 & 27 & 56 & 26 & \un{38} & 35 \\
    && MD\textsubscript{total} && 0 & 0 & 0 & 0 & 0 & 0 & 0 & 0 \\
    && MD\textsubscript{G} && 0 & 0 & 0 & 0 & 0 & 0 & 0 & 0 \\
    && MD\textsubscript{L} && 0 & 0 & 0 & 0 & 0 & 0 & 0 & 0 \\
    && AD && 0 & 0 & 0 & 0 & 0 & 0 & 0& 0 \\
    && Bias && \textbf{2\%} & \textbf{2\%} & \un{6\%} & 3\% & \un{6\%} & 3\% & 4\% & 4\% \\
    \arrayrulecolor{black!30}
    \cmidrule{3-12}
    \arrayrulecolor{black}
    &\multirow{10}{*}{\rotatebox{90}{\textbf{Budget per top.: 300 (17\%)}}} &\# rels. && \un{2042} & 2923 & \textbf{3628} & 2913 & 3607 & 2868 & 3609 & 2723 \\
    && AA && \un{890} & 904 & \textbf{909} & \textbf{909} & \textbf{909} & \textbf{909} & 907 & 906 \\
    && MA\textsubscript{total} && \un{40} & 29 & 30 & \textbf{20} & 31 & 21 & 36 & 32 \\
    && MA\textsubscript{G} && \un{27} & 13 & \textbf{8} & \textbf{8} & \textbf{8} & \textbf{8} & 10 & 11 \\
    && MA\textsubscript{L} && 13 & 16 & 22 & \textbf{12} & 22 & 13 & \un{26}& 21\\
    && MD\textsubscript{total} && 0 & 0 & 0 & 0 & 0 & 0 & 0 & 0\\
    && MD\textsubscript{G} && 0 & 0 & 0 & 0 & 0 & 0 & 0 & 0\\
    && MD\textsubscript{L} && 0 & 0 & 0 & 0 & 0 & 0 & 0 & 0\\
    && AD && 0 & 0 & 0 & 0 & 0 & 0 & 0 & 0\\
    && Bias && \textbf{1\%} & 2\% & 2\% & \textbf{1\%} & 2\% & \textbf{1\%} & \un{3\%} & 2\% \\
    \bottomrule
  \end{tabular}
\end{table}

To support a more detailed analysis, in \Cref{tab:agr-trec8-pooled}, we report the raw agreements of each method.
The upper half of the table includes the results obtained when using AP for evaluating the runs.
In this case, there are a total of 966 gold significant differences ($\lvert T_G \rvert = 966$).
The lower half includes the results when using NDCG.
In this case, there are a total of 917 gold significant differences ($\lvert T_G \rvert = 917$).

The AA counts confirm that adjudication methods are more effective than top-$k$ and NTCIR pooling methods in detecting significant pairs in the correct order,
especially at lower budgets.
They provide further insights about the (almost) 1.00 Recall (see \Cref{tab:pyr-trec8-pooled}) we observed for most adjudication methods.
Indeed, with AP, the gold qrels detect 966 significantly different pairs and the AA counts is (almost) 966,
indicating that the 1.00 Recall is due to significant pairs in the correct order.
The same happens for NDCG, where we observe that most methods obtain AA values near 917.
In other terms, the slight drop in Kendall's $\tau$ observed in \Cref{tab:pyr-trec8-pooled} is not caused by wrongly ordered pairs, even when Recall is 1.00.
When it comes to the specific methods, MTF achieves the best AA figures for budgets of 100, 300 when using AP, while under NDCG Hedge works slightly better
with lower budgets and bandit-based methods perform the best with a budget of 300.

If we compare the AA counts with the number of relevant documents found by a method (the \# rels. row),
we observe a somehow unexpected behaviour.
One might think that the more relevant documents found, the more AA increases.
However, for a budget of 100 judgements per topic, Hedge adjudicated 2170 relevant documents, 485 more than
MTF, but the latter one achieves the highest AA with AP; the same happens again for a budget of 300:
MTF is not the best one in terms of relevant documents but it is the best in terms of AA.
We can observe something similar with NDCG: founding more relevant documents does not necessarily mean more AA.
Obviously, having more relevant documents in the pool helps in increasing the number of AA, but these
results showcase that it is not the only factor.
Overall, these observations suggest that not all the relevant documents are equally discriminative in finding significantly different pairs.
Indeed, relevant documents appear at different ranks in the results lists and the same (or even higher)
number of relevant documents may contribute differently to the performance score of a run and, in turn, to the significant differences found.
So far, research has mostly focused on determining the number of topics needed~\cite{Buckley2000,Sakai2016a,Sanderson2005,Voorhees2002,Voorhees2009}
or on identifying the most discriminative subset of topics~\cite{Hauff2009,Hosseini2012,Mizzaro2007,Roitero2020a}.
These findings open up the possibility of future research on which are the best relevant documents to more reliably discriminate among systems,
an area not well explored yet, to the best of our knowledge.

Almost in every case, no method fails in a mixed or active disagreement,
i.e. detecting significant differences when there is a swap.
This represents a very important insight from this experiment,
since it shows that no method causes a ranking swap between a pair of systems that were originally significantly different.
In other terms, the drop in Kendall's $\tau$ is not due to swaps between systems that are significantly different on the gold
qrels but swaps only happen among not significantly different systems, having a much lower impact.

Let us now consider MA\textsubscript{G} and MA\textsubscript{L}.
The former accounts for significant pairs in the gold qrels which are missed by reduced pools; thus, it helps mainly to explain drops in Recall.
The latter accounts for significant pairs in a reduced pool which are not present in the gold qrels; thus, it helps mainly to explain drops in Precision.
We can observe that MA\textsubscript{G} gets reduced as the budget size increases up to almost 0, with the exception of top-$k$ pooling, Hedge and NTCIR method,
consistently with the previous findings in \Cref{tab:pyr-trec8-pooled}.
Moreover, MA\textsubscript{L} is consistently higher than MA\textsubscript{G},
explaining the loss in Precision even at very high Recall levels.

When it comes to publication bias, we observe moderate values, from 7\% and below,
suggesting that all the methods would not lead to draw conclusions severely different from the gold qrels.
We can observe that bias quickly decreases as the budget increases and that adjudication methods are more effective than top-$k$ pooling,
achieving a bias up to 2-3 times lower than it.

Finally, we can observe that there are not different trends between the two evaluation metrics employed, AP and NDCG.
This shows that the results presented here are not an artefact of the metric used, but of the adjudication methods being evaluated.

% !TeX spellcheck = en_GB

\begin{table}
    \centering
    \footnotesize
    \ra{1.1}
    \caption{Kendall's $\tau$, Precision and Recall (see \Cref{sec:method}) of each
    adjudication method for a varying number of judgements per topic. 10 and 30 are the budget of judgements per topic.
    Parentheses indicate the size with respect to the full pool.
    We used the 66 pooled systems from DL21. For each column, the best values are \textbf{bolded} and the worst ones are \un{underlined}.}
    \label{tab:pyr-dl21-all-depth-10}
    \setlength\tabcolsep{2pt}
    \begin{tabular}{@{}
        l@{\hspace{1\tabcolsep}}
        r@{\hspace{0.8\tabcolsep}}
        r@{\hspace{0.8\tabcolsep}}
        r@{\hspace{1\tabcolsep}}
        c@{\hspace{1\tabcolsep}}
        r@{\hspace{0.8\tabcolsep}}
        r@{\hspace{0.8\tabcolsep}}
        r@{\hspace{1\tabcolsep}}
        c@{\hspace{1\tabcolsep}}
        r@{\hspace{0.8\tabcolsep}}
        r@{\hspace{0.8\tabcolsep}}
        r@{\hspace{1\tabcolsep}}
        c@{\hspace{1\tabcolsep}}
        r@{\hspace{0.8\tabcolsep}}
        r@{\hspace{0.8\tabcolsep}}
        r@{}}
        \toprule
        \multirow{2.3}{*}{\textbf{Method}} &\multicolumn{3}{c}{\textbf{MAP/10} (9\%)}&&\multicolumn{3}{c}{\textbf{MAP/30} (26\%)}&&\multicolumn{3}{c}{\textbf{NDCG/10} (9\%)}&&\multicolumn{3}{c}{\textbf{NDCG/30} (26\%)}\\
        \cmidrule{2-4}\cmidrule{6-8}\cmidrule{10-12}\cmidrule{14-16}
        & \multicolumn{1}{c}{\textbf{$\tau$}} & \multicolumn{1}{c}{\textbf{P}} & \multicolumn{1}{c}{\textbf{R}} &&
        \multicolumn{1}{c}{\textbf{$\tau$}} & \multicolumn{1}{c}{\textbf{P}} & \multicolumn{1}{c}{\textbf{R}} &&
        \multicolumn{1}{c}{\textbf{$\tau$}} & \multicolumn{1}{c}{\textbf{P}} & \multicolumn{1}{c}{\textbf{R}} &&
        \multicolumn{1}{c}{\textbf{$\tau$}} & \multicolumn{1}{c}{\textbf{P}} & \multicolumn{1}{c}{\textbf{R}} \\
        \midrule
        top-$k$    & 0.46 & 0.448 & 0.445 && 0.69 & 0.668 & 0.833 && 0.61 & 0.531 & 0.554 && \textbf{0.82} & 0.723 & 0.832 \\
        MTF        & 0.49 & \textbf{0.611} & \un{0.414} && 0.69 & 0.687 & 0.798 && 0.61 & \textbf{0.632} & 0.534 && 0.79 & 0.734 & 0.808 \\
        MM         & \textbf{0.53} & 0.566 & 0.477 && \textbf{0.73} & \textbf{0.764} & 0.778 && \textbf{0.66} & 0.628 & 0.598 && 0.81 & 0.772 & 0.808 \\
        MM-NS      & 0.50 & 0.517 & 0.505 && 0.70 & 0.654 & 0.841 && 0.64 & 0.593 & 0.607 && \textbf{0.82} & 0.725 & \textbf{0.844} \\
        TS         & 0.52 & 0.554 & 0.489 && \textbf{0.73} & 0.761 & 0.777 && \textbf{0.66} & 0.624 & 0.605 && \textbf{0.82} & \textbf{0.780} & 0.809 \\
        TS-NS      & 0.50 & 0.509 & 0.502 && 0.69 & 0.642 & 0.839 &&  0.63 & 0.589 & 0.603 && 0.81 & 0.715 & 0.839 \\
        Hedge      & \un{0.42} & 0.430 & 0.419 && \un{0.50} & \un{0.558} & \un{0.603} && \un{0.51} & \un{0.521} & \un{0.484} && \un{0.61} & \un{0.657} & \un{0.674} \\
        NTCIR      & 0.47 & \un{0.423} & \textbf{0.560} && 0.69 & 0.594 & \textbf{0.871} && 0.59 & 0.522 & \textbf{0.621} && 0.76 & 0.669 & 0.827 \\
        \bottomrule
    \end{tabular}
\end{table}
% !TeX spellcheck = en_GB

\begin{table}
  \centering
  \footnotesize
  \caption{Relevants, agreements and bias of each adjudication method for a varying
  number of judgements per topic. Parentheses indicate the size with respect to the full pool.
  We used the 66 pooled systems from DL21.
  The top-10 pool includes 3541 relevant documents.
  There are a total of 2145 pairwise comparisons, of which 418 are significant under the gold qrels with MAP (upper half), and 417 with NDCG (lower half).
  For each budget, the best values are \textbf{bolded} and the worst ones are \un{underlined}.}
  \label{tab:agr-dl21-all-depth-10}
  \setlength\tabcolsep{2pt}
  \begin{tabular}{@{}llllrrrrrrrr@{}}
    \toprule
    &&\multirow{2.3}{*}{\textbf{Metric}} && \multicolumn{8}{c}{\textbf{Adjudication method}} \\
    \cmidrule{5-12}
    &&&&\multicolumn{1}{c}{top-\textit{k}}&\multicolumn{1}{c}{MTF}&\multicolumn{1}{c}{MM}&\multicolumn{1}{c}{MM-NS}&\multicolumn{1}{c}{TS}&\multicolumn{1}{c}{TS-NS}&\multicolumn{1}{c}{Hedge}&\multicolumn{1}{c}{NTCIR}\\
    \midrule
    \multirow{20.5}{*}{\rotatebox{90}{\textbf{MAP (418 gold significantly different pairs)}}}&\multirow{10}{*}{\rotatebox{90}{\textbf{Budget per topic: 10 (9\%)}}} &\# rels. && \un{441} & 488 & 489 & 474 & 483 & 469 & 504 & \textbf{513} \\
    && AA && 186 & \un{173} & 199 & 211 & 204 & 210 & 175 & \textbf{234}  \\
    && MA\textsubscript{total} && 413 & \textbf{345} & 358 & 386 & 361 & 392 & 442 & \un{464}  \\
    && MA\textsubscript{G} && 214 & \un{237} & 212 & 201 & 206 & 203 & 235 & \un{176}  \\
    && MA\textsubscript{L} && 199 & \textbf{108} & 146 & 185 & 155 & 189 & 207 & \un{288}  \\
    && MD\textsubscript{total} && \un{48} & \textbf{13} & 15 & 19 & 18 & 19 & 33 & 39  \\
    && MD\textsubscript{G} && \un{18} & 8 & 7 & \textbf{6} & 8 & \textbf{6} & 8 & 8  \\
    && MD\textsubscript{L} && 30 & \textbf{5} & 9 & 13 & 10 & 14 & 25 & \un{31} \\
    && AD && 0 & 0 & 0 & 0 & 0 & 0 & 0 & 0  \\
    && Bias && 55\% & \textbf{39\%} & 43\% & 48\% & 45\% & 49\% & 57\% & \un{58\%}  \\
    \arrayrulecolor{black!30}
    \cmidrule{3-12}
    \arrayrulecolor{black}
    &\multirow{10}{*}{\rotatebox{90}{\textbf{Budget per top.: 30 (26\%)}}} &\# rels. && \un{1186} & 1327 & \textbf{1359} & 1289 & 1345 & 1267 & 1352 & 1337  \\
    && AA && 348 & 334 & 325 & 352 & 325 & 351 & \un{252} & \textbf{364}  \\
    && MA\textsubscript{total} && 243 & 237 & \textbf{194} & 251 & 196 & 262 & \un{355} & 299  \\
    && MA\textsubscript{G} && 70 & 84 & 93 & 66 & 93 & 67 & \un{161} & \textbf{54}  \\
    && MA\textsubscript{L} && 173 & 152 & \textbf{101} & 185 & 103 & 194 & 194 & \un{245}  \\
    && MD\textsubscript{total} && \textbf{0} & \textbf{0} & \textbf{0} & 1 & \textbf{0} & 1 & \un{11} & 4  \\
    && MD\textsubscript{G} && \textbf{0} & \textbf{0} & \textbf{0} & \textbf{0} & \textbf{0} & \textbf{0} & \un{5} & \textbf{0}  \\
    && MD\textsubscript{L} && \textbf{0} & \textbf{0} & \textbf{0} & 1 & \textbf{0} & 1 & \un{6} & 4  \\
    && AD && 0 & 0 & 0 & 0 & 0 & 0 & 0 & 0   \\
    && Bias && 33\% & 31\% & \textbf{24\%} & 35\% & \textbf{24\%} & 36\% & \un{44\%} & 41\%  \\
    \midrule
    \multirow{20.5}{*}{\rotatebox{90}{\textbf{NDCG (417 gold significantly different pairs)}}}&\multirow{10}{*}{\rotatebox{90}{\textbf{Budget per topic: 10 (9\%)}}} &\# rels. && \un{441} & 488 & 489 & 474 & 483 & 469 & 504 & \textbf{513} \\
    && AA && 231 & 223 & 249 & 253 & 252 & 252 & \un{202} & \textbf{259}  \\
    && MA\textsubscript{total} && 376 & 322 & \textbf{314} & 333 & 315 & 337 & \un{388} & 381  \\
    && MA\textsubscript{G} && 184 & 193 & 167 & 164 & 165 & 165 & \un{215} & \textbf{158}  \\
    && MA\textsubscript{L} && 192 & \textbf{129} & 146 & 170 & 151 & 172 & 173 & \un{223}  \\
    && MD\textsubscript{total} && \un{14} & \textbf{3} & \textbf{3} & 4 & \textbf{3} & 5 & 13 & \un{14}  \\
    && MD\textsubscript{G} && \un{2} & 1 & \textbf{0} & \textbf{0} & \textbf{0} & \textbf{0} & \textbf{0} & 0  \\
    && MD\textsubscript{L} && 12 & \textbf{2} & \textbf{2} & 4 & \textbf{2} & 5 & 13 & \un{14}  \\
    && AD && 0 & 0 & 0 & 0 & 0 & 0 & 0 & 0  \\
    && Bias && 47\% & \textbf{37\%} & \textbf{37\%} & 41\% & 38\% & 41\% & \un{48\%} & \un{48\%}  \\
    \arrayrulecolor{black!30}
    \cmidrule{3-12}
    \arrayrulecolor{black}
    &\multirow{10}{*}{\rotatebox{90}{\textbf{Budget per top.: 30 (26\%)}}} &\# rels. && \un{1186} & 1327 & \textbf{1359} & 1289 & 1345 & 1267 & 1352 & 1337  \\
    && AA && 347 & 337 & 337 & \textbf{352} & 338 & 350 & \un{281} & 345 \\
    && MA\textsubscript{total} && 203 & 203 & 180 & 199 & \textbf{175} & 207 & \un{283} & 243  \\
    && MA\textsubscript{G} && 70 & 80 & 80 & \textbf{65} & 79 & 67 & \un{136} & 72  \\
    && MA\textsubscript{L} && 133 & 122 & 101 & 134 & \textbf{96} & 140 & 147 & \un{171}  \\
    && MD\textsubscript{total} && 0 & 0 & 0 & 0 & 0 & 0 & 0 & 0  \\
    && MD\textsubscript{G} && 0 & 0 & 0 & 0 & 0 & 0 & 0 & 0  \\
    && MD\textsubscript{L} && 0 & 0 & 0 & 0 & 0 & 0 & 0 & 0  \\
    && AD && 0 & 0 & 0 & 0 & 0 & 0 & 0 & 0  \\
    && Bias && 28\% & 27\% & 23\% & 27\% & \textbf{22\%} & 29\% & \un{34\%} & 33\%  \\
    \bottomrule
  \end{tabular}
\end{table}

Additionally, we run experiments on the TREC Deep Learning (DL) track 2021.
We selected this collection as having opposing characteristics to TREC-8.
The DL collection adopts a very shallow pooling at just depth 10,
representing a quite challenging setting for adjudication methods.
We believe that using these two collections helps in supporting the generalizability of the results presented here.
\Cref{tab:pyr-dl21-all-depth-10} reports the Kendall's $\tau$, Precision, and Recall, similarly to \Cref{tab:pyr-trec8-pooled} for TREC-8;
\Cref{tab:agr-dl21-all-depth-10} reports the agreement counts, similarly to \Cref{tab:agr-trec8-pooled} for TREC-8.
In general, we observe quite lower and much more varied performance on DL 2021 than on TREC-8.

Kendall's $\tau$ is generally low for all the methods with both metrics.
In TREC-8, adjudication methods were able to obtain very strong results only with a 17\% of the original budget,
while in this case no method is able to reach that performance even with a 26\%.
One important difference is that, while in TREC-8 top-$k$ and NTCIR method were clearly underperforming with respect to the other methods,
in DL 2021 Hedge clearly achieves the worst performance.

When it comes to the agreements (\Cref{tab:pyr-dl21-all-depth-10}),
a notable difference is that, at low budgets (9\%),
MD appear while they go to (almost) zero for higher budgets.
The MD at 9\% budget indicate that the drop in Kendall's $\tau$ are also due to swaps in the significantly different pairs.
The problem concerns more  MD\textsubscript{L}, i.e. swaps in significant pairs detected by a reduced pool but not the gold qrles,
than MD\textsubscript{G}, i.e. swaps in significant pairs detected by the gold qrels but not a reduced pool.
As a consequence, part of the loss of Precision is due to swaps in the significant pairs
a more severe condition than the one causing the loss of Precision in TREC-8.
This issue impacts more top-$k$ and NTCIR than the adjudication methods but, overall,
low budgets and shallow pools do not lead to reliable enough results.

When it comes to AA, differently from TREC-8,
they struggle to get close to the total number of significantly different pairs on the gold qrels.
As in the TREC-8 case, an increase in the number of relevant documents found does not necessarily lead to an increase in the AA counts.

On a positive side, AD is always 0, also for DL 2021.

When it comes to MA, we observe two different patterns.
Differently from TREC-8, MA\textsubscript{G} is always quite high, motivating the general lack of Recall.
In addition, MA\textsubscript{L} does not substantially decrease as the budget increases, explaining the general lack of Precision.

Publication bias is exceedingly high, especially at low budgets, ranging between 25\% and 50\%.
Overall, these high values shed a negative light on the reliability of the conclusions
you would draw when using these methods under shallow pool conditions.

%%%%%%%%%%%%%%%%%%%%%%%%%%%%%%%%%%%%%%%%%%%%%%%%%
% RQ2
%%%%%%%%%%%%%%%%%%%%%%%%%%%%%%%%%%%%%%%%%%%%%%%%%
\subsection{\Cref{rq2}: How and where the methods fail}

% !TeX spellcheck = en_GB

\begin{figure}%
    \centering%
    \begin{subfigure}{0.20\textwidth}%
      \includegraphics[width=\textwidth]{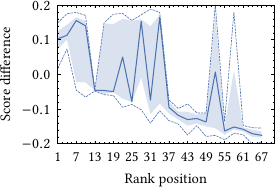}%
      \caption{top-\textit{k}}%
    \end{subfigure}%    
    \begin{subfigure}{0.20\textwidth}%
        \includegraphics[width=\textwidth]{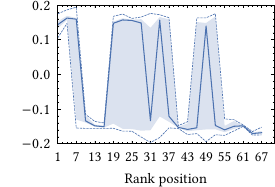}%
        \caption{MTF}%
    \end{subfigure}%

    \vspace*{-2pt}

    \begin{subfigure}{0.20\textwidth}%
        \includegraphics[width=\textwidth]{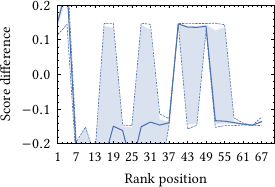}%
        \caption{MM}%
    \end{subfigure}%
     \begin{subfigure}{0.20\textwidth}%
        \includegraphics[width=\textwidth]{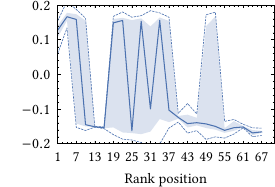}%
        \caption{MM-NS}%
    \end{subfigure}%

    \vspace*{-2pt}

    \begin{subfigure}{0.20\textwidth}%
        \includegraphics[width=\textwidth]{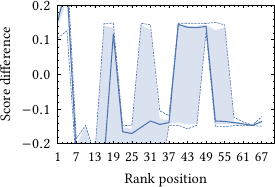}%
        \caption{TS}%
    \end{subfigure}%
    \begin{subfigure}{0.20\textwidth}%
        \includegraphics[width=\textwidth]{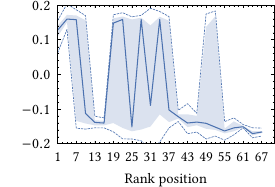}%
        \caption{TS-NS}%
    \end{subfigure}%

    \vspace*{-2pt}
    
    \begin{subfigure}{0.20\textwidth}%
      \includegraphics[width=\textwidth]{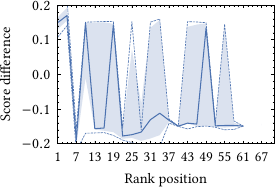}%
      \caption{Hedge}%
    \end{subfigure}%
    \begin{subfigure}{0.20\textwidth}%
      \includegraphics[width=\textwidth]{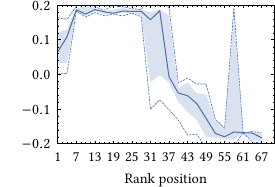}%
      \caption{NTCIR}%
    \end{subfigure}%
    
    \caption{Distribution of MAP differences between systems in MA for a budget of 100 assessments (6\%). 
    The x-axis represents the systems sorted by their position in the official ranking.
    Each data point holds the distribution of 3 systems.
    The solid line represents the median of the bin.
    The shaded area is limited by the first and third quartiles of the distribution, i.e. it represents the inter-quartile range.
    Finally, the dashed lines are the maximum and the minimum.
    Breaks in the lines mean that there was not any mixed agreement for those systems.
    We used the 71 pooled systems of TREC-8.}
    \label{fig:ma-boxplot-trec8-pooled-difference}%
\end{figure}

We study how and where, in terms of rank positions, the different methods fail in detecting significant differences.

We focus our analysis on the cases of mixed agreements (MA),
which have shown to be the main factor for the loss of Precision and Recall.
\Cref{fig:ma-boxplot-trec8-pooled-difference} shows the distribution of the score differences in systems pairs which belong to MA
with respect to their position in the gold ranking of systems for a budget of 100 assessments (6\%).
For each MA pair, we compute the difference between the score of the best and the worst system in the pair (under the adjudicated qrels, not the gold ones),
recording it with a positive sign for the best system and a negative one for the worst system.
\footnote{For example, if we have the pair of system1 and system2 in mixed agreement, and system1 has the highest score, and their score difference is $0.15$ (with the reduced pool).
Then, for system1 we record $0.15$ and for system2 we save $-0.15$. The mentioned figure plots the distribution of these differences for each system, according
to their position in the ranking induced with the gold qrels.}
\Cref{fig:ma-boxplot-trec8-pooled-difference} tries to convey information about the distribution of such differences as a series of boxplots would do,
but in a more compact and reabable way.
The x-axis is the position of each system in the ranking of systems under the gold qrels,
and we consider bins of three rank positions to make the figure more readable.
For example, the first point in the figure represents the distribution of the mentioned differences for the first three systems in the gold ranking of systems.
The solid line represents the median of the bin;
the shaded area is limited by the first and third quartiles of the
distribution, i.e. it represents the inter-quartile range;
finally, the dashed lines are the maximum and the minimum.
A break in the lines means that no pair of systems in that range of rank positions is a MA.

We can see some clear trends among all the evaluated methods.
As a general trend for most adjudication methods,
the biggest differences occur between MA systems in the middle of the ranking
(we see wider areas in the middle of the ranking),
whereas we see more narrow distributions in the top-ranked and lowest-ranked methods.
This suggests that the MA, and the consequent loss of Precision, happen in a region of moderate impact,
since mid-rank systems may receive less interest in any case.
Top-$k$ and NTCIR method represent two notable exceptions.
Indeed, top-$k$ concentrates most of the score differences in the top ranks;
therefore, top-$k$ is not only the less performing method
(see \Cref{tab:pyr-trec8-pooled} and \Cref{tab:agr-trec8-pooled})
but it also fails in the most impactful region of the ranking.
This is even worse for NTCIR, where the biggest differences (of 0.2 points),
are all clustered in the top positions of the ranking.

%%%%%%%%%%%%%%%%%%%%%%%%%%%%%%%%%%%%%%%%%%%%%%%%%
% RQ3
%%%%%%%%%%%%%%%%%%%%%%%%%%%%%%%%%%%%%%%%%%%%%%%%%
\subsection{\Cref{rq3}: Evaluation of unseen systems}

We investigate the \emph{reusability} of the judgements produced by a low-cost method,
i.e. their ability to fairly evaluate unseen systems.
Usually, reusability is evaluated by following a \textit{leave-one-group-out} approach.
This consists in forming pools leaving one participating group each time and using those pools
to evaluate the submissions of the group that was left out.
We follow a different approach using the non-pooled systems of TREC-8.
\footnote{We do not perform these experiments on the DL21 collection since it does not include non-pooled runs.}
To this aim, we performed the same experiments as in the previous sections,
but using the non-pooled systems of TREC-8.
In this way, we are evaluating systems that did not participate in the constructions of the pools.
As commented in \Cref{sec:experiments}, this collection has been repeatedly acknowledged in the community as a high-quality one to evaluate unseen systems.
Thus, we assume that the TREC-8 gold judgements are reusable and,
if a low-cost method provides the same significant differences as them,
we conclude that it is reusable as well.

% !TeX spellcheck = en_GB

\begin{table}
    \centering
    \footnotesize
    \ra{1}
    \caption{Kendall's $\tau$, Precision and Recall (see \Cref{sec:method}) of each
    adjudication method for a varying number of judgements per topic. 
    100 and 300 are the budget of judgements per topic.
    Parentheses indicate the size with respect to the full pool.
    We used the 58 non-pooled systems from TREC-8.
    For each budget, best values are \textbf{bolded} and worst ones \un{underlined}.}
    \setlength\tabcolsep{2pt}
    \label{tab:pyr-trec8-non-pooled}
    \begin{tabular}{@{}
        l@{\hspace{1\tabcolsep}}
        r@{\hspace{0.8\tabcolsep}}
        r@{\hspace{0.8\tabcolsep}}
        r@{\hspace{1\tabcolsep}}
        c@{\hspace{1\tabcolsep}}
        r@{\hspace{0.8\tabcolsep}}
        r@{\hspace{0.8\tabcolsep}}
        r@{\hspace{1\tabcolsep}}
        c@{\hspace{1\tabcolsep}}
        r@{\hspace{0.8\tabcolsep}}
        r@{\hspace{0.8\tabcolsep}}
        r@{\hspace{1\tabcolsep}}
        c@{\hspace{1\tabcolsep}}
        r@{\hspace{0.8\tabcolsep}}
        r@{\hspace{0.8\tabcolsep}}
        r@{}}
      \toprule
      \multirow{2.3}{*}{\textbf{Method}} &\multicolumn{3}{c}{\textbf{MAP/100} (6\%)}&&\multicolumn{3}{c}{\textbf{MAP/300} (17\%)}&&\multicolumn{3}{c}{\textbf{NDCG/100} (6\%)}&&\multicolumn{3}{c}{\textbf{NDCG/300} (17\%)}\\
      \cmidrule{2-4}\cmidrule{6-8}\cmidrule{10-12}\cmidrule{14-16}
      & \multicolumn{1}{c}{\textbf{$\tau$}} & \multicolumn{1}{c}{\textbf{P}} & \multicolumn{1}{c}{\textbf{R}} &&
      \multicolumn{1}{c}{\textbf{$\tau$}} & \multicolumn{1}{c}{\textbf{P}} & \multicolumn{1}{c}{\textbf{R}} &&
      \multicolumn{1}{c}{\textbf{$\tau$}} & \multicolumn{1}{c}{\textbf{P}} & \multicolumn{1}{c}{\textbf{R}} &&
      \multicolumn{1}{c}{\textbf{$\tau$}} & \multicolumn{1}{c}{\textbf{P}} & \multicolumn{1}{c}{\textbf{R}} \\
      \midrule
      top-$k$    & \un{0.82} & 0.931 & \un{0.903} && \un{0.91} & \un{0.948} & \un{0.966} && \un{0.83} & 0.941 & \un{0.880} && \un{0.90} & \un{0.966} & \un{0.943} \\
      MTF        & 0.88 & 0.934 & 0.933 && 0.95 & 0.968 & 0.988 && 0.89 & 0.941 & 0.916 && 0.94 & 0.980 & 0.968 \\
      MM         & \textbf{0.91} & 0.967 & \textbf{0.942} && \textbf{0.97} & 0.976 & \textbf{0.997} && 0.92 & 0.955 & \textbf{0.946} && \textbf{0.97} & \textbf{0.983} & \textbf{0.979} \\
      MM-NS      & 0.88 & 0.948 & 0.936 && 0.96 & 0.966 & 0.989 &&  0.88 & 0.952 & 0.921 && 0.95 & 0.978 & 0.976 \\
      TS         & \textbf{0.91} & 0.969 & 0.940 && \textbf{0.97} & 0.973 & 0.996 && 0.92 & 0.956 & 0.944 && \textbf{0.97} & 0.979 & 0.977 \\
      TS-NS      & 0.87 & 0.945 & 0.933 && 0.95 & 0.966 & 0.986 && 0.88 & 0.952 & 0.918 && 0.94 & 0.979 & 0.974 \\
      Hedge      & \textbf{0.91} & \textbf{0.973} & 0.929 && 0.96 & \textbf{0.980} & 0.982 && \textbf{0.93} & \textbf{0.974} & \textbf{0.946} && 0.96 & 0.977 & 0.977 \\
      NTCIR      & 0.89 & \un{0.898} & 0.931 && 0.95 & 0.962 & 0.984 && 0.86 & \un{0.938} & 0.911 && 0.94 & 0.974 & 0.977 \\
      \bottomrule
    \end{tabular}
\end{table}
% !TeX spellcheck = en_GB

\begin{table}
  \centering
  \footnotesize
  \caption{Relevants, agreements and bias of each adjudication method for a varying
  number of judgements per topic. Parentheses indicate the size with respect to the full pool.
  We used the 58 non-pooled systems from TREC-8.
  The top-100 full pool includes 4728 relevant documents.
  There are 1653 pairwise comparisons, of which 509 are significant under the gold qrels with MAP (upper half), and 527 with NDCG (lower half).
  For each budget, the best values are \textbf{bolded} and the worst ones are \un{underlined}.}
  \label{tab:agr-trec8-nonpooled}
  \setlength\tabcolsep{2pt}
  \begin{tabular}{@{}llllrrrrrrrr@{}}
    \toprule
    &&\multirow{2.3}{*}{\textbf{Metric}} && \multicolumn{8}{c}{\textbf{Adjudication method}} \\
    \cmidrule{5-12}
    &&&&\multicolumn{1}{c}{top-\textit{k}}&\multicolumn{1}{c}{MTF}&\multicolumn{1}{c}{MM}&\multicolumn{1}{c}{MM-NS}&\multicolumn{1}{c}{TS}&\multicolumn{1}{c}{TS-NS}&\multicolumn{1}{c}{Hedge}&\multicolumn{1}{c}{NTCIR}\\
    \midrule
    \multirow{20.5}{*}{\rotatebox{90}{\textbf{MAP (509 gold significantly different pairs)}}}&\multirow{10}{*}{\rotatebox{90}{\textbf{Budget per topic: 100 (6\%)}}} &\# rels. && \un{1077} & 1685 & 2148 & 1553 & 2102 & 1514 & \textbf{2170} & 1481 \\
    && AA && \un{460} & 475 & \textbf{480} & 477 & 479 & 475 & 473 & 474\\
    && MA\textsubscript{total} && 83 & 68 & \textbf{45} & 58 & \textbf{45} & 62 & 49 & \un{89}\\
    && MA\textsubscript{G} && \un{49} & 34 & \textbf{29} & 32 & 30 & 34 & 36 & 35\\
    && MA\textsubscript{L} && 34 & 34 & 16 & 26 & 15 & 28 & \textbf{13} & \un{54}\\
    && MD\textsubscript{total} && 0 & 0 & 0 & 0 & 0 & 0 & 0 & 0\\
    && MD\textsubscript{G} && 0 & 0 & 0 & 0 & 0 & 0 & 0 & 0\\
    && MD\textsubscript{L} && 0 & 0 & 0 & 0 & 0 & 0 & 0 & 0\\
    && AD && 0 & 0 & 0 & 0 & 0 & 0 & 0 & 0\\
    && Bias && \textbf{7\%} & \textbf{7\%} & \un{3\%} & 5\% & \un{3\%} & 5\% & \un{3\%} & 4\%\\
    \arrayrulecolor{black!30}
    \cmidrule{3-12}
    \arrayrulecolor{black}
    &\multirow{10}{*}{\rotatebox{90}{\textbf{Budget per top.: 300 (17\%)}}} &\# rels. && \un{2042} & 2923 & \textbf{3628} & 2913 & 3607 & 2868 & 3609 & 2723\\
    && AA && \un{492} & 503 & \textbf{508} & 504 & 507 & 502 & 500 & 501\\
    && MA\textsubscript{total} && \un{44} & 23 & \textbf{13} & 23 & 16 & 25 & 19 & 28\\
    && MA\textsubscript{G} && \un{17} & 6 & \textbf{1} & 5 & 2 & 7 & 9 & 8\\
    && MA\textsubscript{L} && \un{27} & 17 & 12 & 18 & 14 & 18 & \textbf{10} & 20\\
    && MD\textsubscript{total} && 0 & 0 & 0 & 0 & 0 & 0 & 0 & 0\\
    && MD\textsubscript{G} && 0 & 0 & 0 & 0 & 0 & 0 & 0 & 0\\
    && MD\textsubscript{L} && 0 & 0 & 0 & 0 & 0 & 0 & 0 & 0\\
    && AD && 0 & 0 & 0 & 0 & 0 & 0 & 0 & 0\\
    && Bias && \un{5\%} & 3\% & \textbf{2\%} & 3\% & 3\% & 3\% & \textbf{2\%} & 4\%\\
    \midrule
    \multirow{20.5}{*}{\rotatebox{90}{\textbf{NDCG (527 gold significantly different pairs)}}}&\multirow{10}{*}{\rotatebox{90}{\textbf{Budget per topic: 100 (6\%)}}} &\# rels. && \un{1077} & 1685 & 2148 & 1553 & 2102 & 1514 & \textbf{2170} & 1481\\
    && AA && \un{464} & 483 & \textbf{499} & 486 & 498 & 484 & \textbf{499} & 480\\
    && MA\textsubscript{total} && \un{92} & 74 & 52 & 65 & 52 & 67 & \textbf{41}& 79 \\
    && MA\textsubscript{G} && \un{63} & 44 & \textbf{28} & 41 & 29 & 43 & \textbf{28} & 47\\
    && MA\textsubscript{L} && 29 & 30 & 23 & 24 & 23 & 24 & \textbf{13} & \un{32}\\
    && MD\textsubscript{total} && 0 & 0 & 0 & 0 & 0 & 0 & 0 & 0\\
    && MD\textsubscript{G} && 0 & 0 & 0 & 0 & 0 & 0 & 0 & 0\\
    && MD\textsubscript{L} && 0 & 0 & 0 & 0 & 0 & 0 & 0 & 0\\
    && AD && 0 & 0 & 0 & 0 & 0 & 0 & 0 & 0\\
    && Bias && \un{6\%} & \un{6\%} & 4\% & 5\% & 4\% & 5\% & \textbf{3\%} & \un{6\%}\\
    \arrayrulecolor{black!30}
    \cmidrule{3-12}
    \arrayrulecolor{black}
    &\multirow{10}{*}{\rotatebox{90}{\textbf{Budget per top.: 300 (17\%)}}} &\# rels. && \un{2042} & 2923 & \textbf{3628} & 2913 & 3607 & 2868 & 3609 & 2723\\
    && AA && \un{497} & 510 & \textbf{516} & 514 & 515 & 514 & 515 & 515\\
    && MA\textsubscript{total} && \un{47} & 27 & \textbf{19} & 24 & 23 & 24 & 24 & 26\\
    && MA\textsubscript{G} && \un{30} & 17 & \textbf{11} & 13 & 12 & 13 & 12& 12 \\
    && MA\textsubscript{L} && \un{17} & 10 & \textbf{9} & 11 & 11 & 11 & 12 & 14\\
    && MD\textsubscript{total} && 0 & 0 & 0 & 0 & 0 & 0 & 0& 0 \\
    && MD\textsubscript{G} && 0 & 0 & 0 & 0 & 0 & 0 & 0 & 0\\
    && MD\textsubscript{L} && 0 & 0 & 0 & 0 & 0 & 0 & 0 & 0\\
    && AD && 0 & 0 & 0 & 0 & 0 & 0 & 0  & 0\\
    && Bias && \un{3\%} & \textbf{2\%} & \textbf{2\%} & \textbf{2\%} & \textbf{2\%} & \textbf{2\%} & \textbf{2\%} & \un{3\%}\\
    \bottomrule
  \end{tabular}
\end{table}

\Cref{tab:pyr-trec8-non-pooled} reports the Kendall's $\tau$, Precision and Recall values of every method,
for a varying number of assessments per topic, using the non-pooled systems.
On a positive side, \Cref{tab:pyr-trec8-non-pooled} shows similar trends as \Cref{tab:pyr-trec8-pooled},
suggesting that there is not a specific bias against non-pooled systems.
On a slightly negative side, we observe that performance in \Cref{tab:pyr-trec8-non-pooled}
are generally slightly lower than those in \Cref{tab:pyr-trec8-pooled}, especially at the lowest budget,
indicating a bit more loss and some more swaps due to not being pooled.

More in detail, TS, MM and Hedge always have the highest correlation scores and while MM achieves always the best Recall,
independently from the budget and the metric.
This means that if we were to gather the judgements of a new collection,
MM would be the best option in terms of reusability of the collected assessments.
As before, top-\textit{k} and NTCIR method lag behind the other methods in all the cases and for every considered measure.
This finding suggests that other alternative methods might be a better option to gather assessments when constructing new experimental collections.

\Cref{tab:agr-trec8-nonpooled} reports the agreements for the non-pooled systems,
similarly to \Cref{tab:agr-trec8-pooled} for the pooled ones.
\footnote{Note that the \# rels. row is the same as before since the pools are the same, we are only changing the systems we are evaluating.}
The results follow the same trends as with the pooled systems, further supporting the lack of strong biases against non-pooled systems.
These scores confirm that alternative adjudication methods are more effective than top-$k$,
which, contrary to what we observed in \Cref{tab:agr-trec8-pooled}, now is clearly the worst method.
As before, the more relevant documents found does not necessarily mean the more AA;
therefore, not all the relevant documents are equally discriminative also for non-pooled systems.

No method fails in a mixed or active disagreement when evaluating the non-pooled systems.
This further supports the fact that most drops in Kendall's $\tau$ are due to swaps
between systems that are not significantly different under the gold qrels.

When it comes to the publication bias, we observe similar trends as in the case of the pooled systems,
even with lower values, indicating that published conclusions would not change also in the case of non-pooled systems.

Finally, we can observe similar trends between the results obtained with AP and those obtained with NDCG,
supporting the fact that the results presented here are generalizable in terms of the evaluation of unseen systems,
and that they are not an artefact of the evaluation metric used.
% !TeX spellcheck = en_GB

\section{Conclusions and Future Work}
\label{sec:fw}

We argued for the need of a more powerful way of evaluating adjudication methods.
In particular, while the current approach just focuses on how close two alternative methods rank systems,
quantified by Kendall's $\tau$, we think that we should focus our attention also on how different methods behave
with respect to the significantly different pairs of systems detected.
Indeed, while the current approach looks for stability in answering the question ``is system A better than B?'',
our proposed method looks for stability in answering the question ``is system A significantly better than B?'',
which is the ultimate questions researchers are interested in to ensure generalizability of results.

To this end, we considered two measures---namely Precision and Recall---which consider significantly different pairs in isolation,
as well as measures---the agreement/disagreement counts---which relate them to swaps in the ranking of systems.
We also considered the problem of the publication bias,
i.e. the chance of publishing results/conclusions that would not hold or be the opposite when using the full pool instead of a reduced one.

To both validate and to showcase our proposed approach, we conducted a thorough experimentation on TREC-8,
a collection renown for its high quality deep pool, and TREC Deep Learning 2021, a collection adopting a very shallow pool.
In this way, we have shown that our methodology allows us to obtain insights not possible simply using Kendall's $\tau$.

For example, we found that no active disagreements (AD) and (almost) no mixed disagreements (MD) happen.
This means that observed drops in Kendall's $\tau$ are mostly due to swaps between not significantly different systems.
Therefore, those drops concerns not very interesting system pairs,
and it might not be worth to strive for (or to judge a method just by) 1.00 Kendall's $\tau$.

We also found that the number of relevant documents detected by a method does not necessarily
increase the number of significantly different pairs detected,
suggesting that not all the relevant documents in a pool are equally discriminative.
This opens up interesting future investigations on which (relevant) documents would be optimal for a
pool while the current focus has been more on determining how many and which topics to sample.

We have shown that drops in Precision and Recall are caused by mixed agreements (MA)
which distribute unevenly at different rank positions and, therefore,
they have a quite different impact:
those happening at mid-to-bottom rank positions are less serious than those happening at the top positions of the ranking.

Finally, we also found that no adjudication methods induces strong biases against non-pooled systems,
thus further supporting the use of these methods to construct new test collections for IR evaluation.
Previous work evaluated the reusability of bandit-based methods using Kendall's $\tau$ and other swap-based measures,
and concluded that the collections built with them were less reusable than desirable.
With the new evaluation approach we have presented in this paper, we shed some more light on this issue and show that,
when focusing on significance between systems, bandit-based method are indeed reusable.

Overall, our approach allowed us to show that existing methods for human assessment
adjudication in IR evaluation could preserve most of the true statistical differences between the
pairwise comparisons of systems.
Besides this, as discussed in detail,
our approach allowed us to pinpoint which adjudication method works better in specific conditions,
why, and how it is different from other methods.
This will thus be a helpful tool and guidance for researchers, when they have to decide which method to choose in their settings.

%%%%%%%%%%%%%%%%%%%%%%%%%%%%%%%%%%%%%%%%%%%%%%%%%
% Acknowledgements
%%%%%%%%%%%%%%%%%%%%%%%%%%%%%%%%%%%%%%%%%%%%%%%%%
\begin{acks}
  This work has received support from:
  \begin{enumerate*}[label=(\roman*)]
    \item project PLEC2021-007662 (MCIN/AEI/10.13039/501100011033, Ministerio de Ciencia e Innovación, Agencia Estatal de Investigación, Plan de Recuperación, Transformación y Resiliencia, Unión Europea-Next GenerationEU);
    \item Programa de Ayudas para la Formación de Profesorado Universitario, grant number FPU20/02659 (Ministerio de Universidades);
    \item project PID2022-137061OB-C21 (Proyectos de Generación de Conocimiento, MCIN);
    \item project ED431-B 2022/33 (Xunta de Galicia/ERDF);
    \item CAMEO, PRIN 2022 n. 2022ZLL7MW.
  \end{enumerate*}
\end{acks}

%%%%%%%%%%%%%%%%%%%%%%%%%%%%%%%%%%%%%%%%%%%%%%%%%
% References
%%%%%%%%%%%%%%%%%%%%%%%%%%%%%%%%%%%%%%%%%%%%%%%%%
\printbibliography

\end{document}